\documentclass[prb,preprint,showpacs]{revtex4}
\usepackage{amsmath}
\usepackage{bbm}
\usepackage{graphicx}

\begin{document}

\def\fin{\hbox{${\vcenter{\vbox{            
   \hrule height 0.4pt\hbox{\vrule width 0.4pt height 6pt
   \kern5pt\vrule width 0.4pt}\hrule height 0.4pt}}}$}}

\title{Periodized discrete elasticity models for defects in graphene}
\author{A. Carpio\cite{carpio:email}} 
\affiliation{ Departamento de Matem\'atica Aplicada,
Universidad Complutense de Madrid; 28040 Madrid, Spain}
\author{L.L. Bonilla\cite{bonilla:email}} 
\affiliation{G. Mill\'an Institute for Fluid Dynamics, Nanoscience and Industrial 
Mathematics, Universidad Carlos III de Madrid; 28911 Legan\'es, Spain}
\date{\today}

\begin{abstract}
The cores of edge dislocations, edge dislocation dipoles and edge dislocation loops in
planar graphene have been studied by means of periodized discrete elasticity models. To 
build these models, we have found a way to discretize linear elasticity on a planar hexagonal 
lattice using combinations of difference operators that do not involve symmetrically all the 
neighbors of an atom. At zero temperature, dynamically stable cores of edge dislocations may
be heptagon-pentagon pairs (glide dislocations) or octagons (shuffle dislocations) depending 
on the choice of initial configuration. Possible cores of edge dislocation dipoles are vacancies, 
pentagon-octagon-pentagon divacancies, Stone-Wales defects and 7-5-5-7 defects. While 
symmetric vacancies, divacancies and 7-5-5-7 defects are dynamically stable, asymmetric 
vacancies and 5-7-7-5 Stone-Wales defects seem to be unstable. 
\end{abstract} 


\maketitle

\section{Introduction}
\label{sec:intro}
Graphene \cite{nov04,mey07} and other two dimensional (2D) crystals \cite{nov05} have 
been experimentally observed quite recently. This discovery has led to new physics where
quantum relativistic phenomena can be mimicked and tested experimentally in condensed
matter physics \cite{gei07}. Among striking electronic properties of graphene due to its
quantum electrodynamics-like spectrum, there are two chiral quantum Hall effects, minimum
quantum conductivity in the limit of vanishing concentration of charge carriers and strong
suppression of quantum interference effects. Ballistic transport on submicron distances at 
room temperature makes graphene a promising material for nanoelectronics  \cite{gei07}.

Defects in graphene strongly affect its electronic and magnetic properties \cite{rotos,per06,rmp}, 
which may be described by the Dirac equation on curved space \cite{cor07}. 
Irradiation experiments show that pentagon-heptagon pairs (5-7 defects), vacancies, 
divacancies (5-8-5 defects comprising an octagon and two adjacent pentagons) and adatoms 
are commonly obtained, but seemingly not Stone-Wales (SW) defects (two adjacent 5-7 
defects with opposite Burgers vectors and whose heptagons share one side, briefly 5-7-7-5 
defects) \cite{experimentos}. The far field of 5-7 defects corresponds to that of 
edge dislocations in elasticity, while the far field of vacancies and divacancies is that of 
an edge dislocation dipole. Quite recently, experimental observations of edge dislocations on 
high-quality graphene grown on Ir(111) have been reported \cite{CNBM}. Studies of 
defects and their motion are important to assess the mechanical response of graphene at the 
atomic scale and, as indicated above, its electronic properties. 

One common way to describe defects in graphene is to use ab initio calculations. Density 
functional theory (DFT) has been used to ascertain the magnetic properties of graphene 
sheets and single-wall nanotubes with vacancies \cite{rotos}. Local spin DFT has been
used to describe glide and shuffle dislocations in irradiated graphitic structures 
\cite{EHB02}. Molecular dynamics (MD) has been used to discuss the stability of nanotubes
under tension \cite{yak96} and also in the presence of different 5-7 pairs such as 5-7-7-5 
and 7-5-5-7 (similar to SW but now the pentagons share one side) defects \cite{orl99}. 
Atomistic Monte Carlo simulations are less costly than MD and have been used in studies of 
the stability of single graphene sheets \cite{fas07}. Classical structural models of graphene 
may account for chirality effects in nanotubes and allow to assess the impact of the lattice 
structure on some elastic properties \cite{cs}. However, these classical models lack the 
ability to generate and move defects. The approach presented in this paper is different. 

Relevant defects in graphene are the cores of different edge dislocation and edge dislocation 
dipoles. Thus we will regularize appropriately linear elasticity on the graphene honeycomb
lattice and describe which are the stable cores of different edge dislocations and dipoles
\cite{stable}. It is well known that cores of dislocations in crystals with covalent bonds are 
very narrow, so that the elastic field decays quite fast to that given by linear elasticity as the 
distance to the dislocation point (in 2D the dislocation line is a point) increases 
\cite{hirth,hull}. This means that we can regularize 
linear elasticity on a relatively small hexagonal lattice and impose boundary conditions 
corresponding to the elastic field of an edge dislocation (or a dislocation dipole) on a boundary 
which is sufficiently far from the dislocation point for the differences of the displacement 
vector to be well approximated by their corresponding differentials. The result will match 
seamlessly a calculation on a much larger lattice provided the far field of a dislocation is the 
same as that given by linear elasticity as we depart from the dislocation point. Despite the slow 
decay of the elastic strain away from the dislocation point, differences and differentials of the 
elastic displacement become indistinguishable a few atoms away from the dislocation point.

Recently, we have developed periodized discrete elasticity models of dislocations
in cubic crystals that describe their motion and interaction  \cite{dm1,dm2,pcb}. These 
models appear to provide the simplest correction to the equations of elasticity allowing 
nucleation and motion of defects and have two main ingredients. Firstly, by discretizing 
elasticity, a linear lattice model involving nearest and next-nearest neighbors is found. In the 
continuum limit, this lattice model provides the equations of elasticity with the appropiate 
crystal symmetry. Secondly, we need to account for the fact that certain atoms at the core
of a dislocation change their next neighbors as the dislocation moves. Thus to describe 
dislocation motion we need an algorithm to update neighbors. Alternatively, we can
periodize discrete differences along the primitive directions of the crystal by using an 
appropriate periodic function, thereby restoring crystal periodicity. This periodic function is 
selected in such a way that elasticity equations are recovered far from defect cores and stable 
static defects can be generated using their known elastic far fields
at zero applied stress. For applied stresses surpassing the Peierls stress of the material, the 
defects should move and the value of the Peierls stress can be used to calibrate the 
periodic function \cite{car03}.

Here we extend this idea to the study  of defects and their impact on the mechanical response 
of graphene sheets at zero temperature. Mostly planar graphene is very different from 
cylindrical carbon nanotubes having very small radii and large curvature and from 3D 
graphite whose descriptions necessarily require studies different from the present one. Two 
facts complicate the task of designing lattice 
models of graphene. One is that graphene has a two-atom basis. The other is that a planar 
hexagonal lattice is intrinsically anisotropic even if its continuum limit is isotropic elasticity.
We have dealt with these issues by using difference operators whose continuum limits 
are linearly independent combinations of the partial derivatives entering the Navier 
equations of linear elasticity. The key idea to choose our difference operators is that they do 
not have to involve all neighbors of an atom on an equal footing. Once we have an 
appropriate discretization of linear elasticity, we periodize the resulting lattice model in a 
way that allows dislocation gliding. Adding thermal effects and local curvature effects due to 
ripples \cite{mey07,fas07} 
increases the complexity of models and we have omitted these effects in the present paper,
which is organized as follows. Section \ref{sec:graphene} recalls some basic details on the 
structure of graphene lattices and presents the stable defects we have obtained by solving the 
periodized discrete elasticity models. The basis of these models are lattice models obtained by 
discretizing elasticity, as indicated in Section \ref{sec:modgraf}. Periodized discrete 
elasticity is discussed in Section \ref{sec:periodic}. Section \ref{sec:defgraf} describes 
numerical tests of defects in graphene carried out with the full 2D model and with a scalar 
reduced version thereof. Among solutions of our models, we have found a stable octagon 
defect. Known defects such as pentagon-heptagon pairs and 5-7-7-5 SW defects move and 
interact as expected. In particular, the two 5-7 defects comprising the two edge dislocations 
with opposite Burgers vectors of a SW glide to each other on their common gliding line
and annihilate. This is not the case if we have a 7-5-5-7 defect (similar to the SW, but now
the pentagons share a common side) \cite{orl99}, which is stable because the dislocation 
centers of the component edge dislocations are displaced one atomic distance from each other 
in different glide lines \cite{geli}. As observed in experiments, we have also obtained stable 
5-8-5 divacancies \cite{experimentos} and symmetric vacancies \cite{kel98}. We find 
that a symmetry-breaking vacancy \cite{nor03} evolves toward a simple threefold symmetric 
vacancy \cite{kel98}. Lastly, Section \ref{sec:conclusions} contains our conclusions.

\section{Structural characteristics of graphene and defects}
\label{sec:graphene}

\begin{figure}
\begin{center}
\includegraphics[width=8cm]{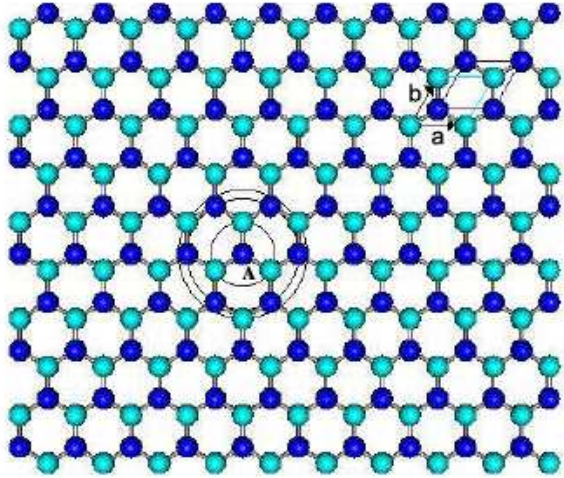}
\caption{(Color online) Graphene hexagonal lattice.  
The unit cell vectors are
${\bf a}=a(1,0)$ and ${\bf b}=a(\cos({\pi\over 3}),\sin({\pi\over 3}))$, 
with  a lattice constant $a=2.461$ angstroms. The unit cell contains two 
carbon atoms $A=(0,0)$ and $B=a(\cos({\pi\over 6}),\sin({\pi\over 6}))$ 
belonging to  two sublattices. An atom $A$ has three nearest neighbors, 
six next-nearest neighbors and three second-nearest neighbors.}
\label{figura1}
\end{center}
\end{figure}

Figure \ref{figura1} illustrates the structure of a {\em graphene sheet} comprising a 
single-layer of graphite. This 2D hexagonal lattice is equivalent to 
a cubic lattice with a two-atom basis generated by two non orthogonal unit cell vectors, 
$\mathbf{a}= (1,0) a$ and $\mathbf{b}= (1,\sqrt{3})a/2$, where $a$ is the
lattice constant. The length of a hexagon side is $l=a/\sqrt{3}$. Dark and light colors 
are used to distinguish the two sublattices: dark atoms belong to sublattice 1 and light ones
to sublattice 2.

\begin{figure}
\begin{center}
\includegraphics[width=8cm]{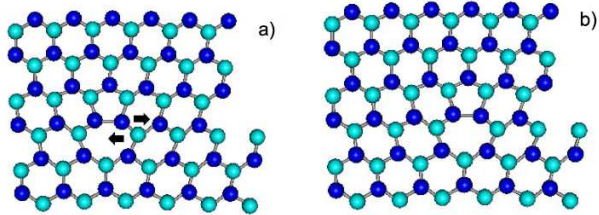}
\caption{(Color online) Numerically generated single pentagon-heptagon defect and its 
motion. The pentagon marks the lower end of an extra column of hexagons and its basis 
forms the top of the heptagon. (a) Under sufficiently large applied shear stress, a dark atom 
moves to the right and a light one to the left as indicated by the arrows. (b) The result is that 
the pentagon-heptagon defect moves one step to the right.}
\label{figura2}
\end{center}
\end{figure}

\begin{figure}
\begin{center}
\includegraphics[width=8cm]{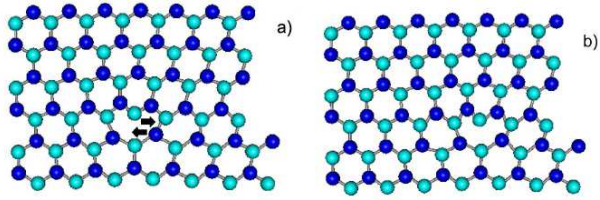}
\caption{(Color online) Numerically generated shuffle dislocation with dangling
bond. An octagon marks the position of the core. (a) Under sufficiently large applied 
shear stress, a light atom moves to the right and a dark one to the left as
indicated by the arrows. (b) The result is that the edge dislocation moves 
one step to the right.}
\label{figura4}
\end{center}
\end{figure}

\begin{figure}
\begin{center}
\includegraphics[width=8cm]{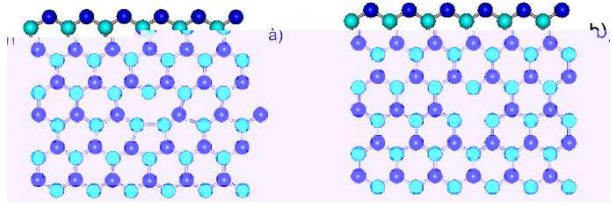}
\caption{(Color online) Numerically generated edge dislocation dipoles.  
Arrangement of atoms in a vacancy: (a) initial configuration as given by linear
elasticity; (b) final configuration after time relaxation in the unstressed lattice.}
\label{figura5}
\end{center}
\end{figure}

\begin{figure}
\begin{center}
\includegraphics[width=8cm]{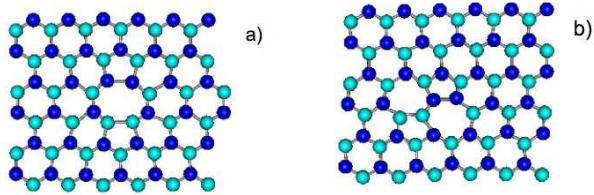}
\caption{(Color online) Numerically generated edge dislocation dipoles. (a) 5-8-5 
divacancy. (b) 5-7-7-5 Stone-Wales defect. }
\label{figura3}
\end{center}
\end{figure}

Graphene layers often contain defects: an experimental study of defects generated by 
irradiation and images thereof is reported in Ref.~\onlinecite{experimentos}. Several 
defects have been obtained by using the periodized discrete vectorial model presented in the 
next Section. They appear in Figures \ref{figura2}, \ref{figura4}, \ref{figura5} and 
\ref{figura3}.  {\em In these figures and successive ones in the paper, only the positions of the atoms
have been calculated numerically. The bonds between neighboring atoms are only aids to 
visualize the structure of the lattice.} 

A simple defect in a graphene sheet is the single pentagon-heptagon (5-7) pair,
depicted in Figures \ref{figura2}(a) and (b). As we will show later, this defect represents an edge dislocation 
and its gliding motion can be characterized by atom 
motion together with breakup and attachment of bonds between atoms.
Note that link breakup and union occurs only between atoms in the direction of defect
motion. As a result of this motion, the neighbors of the moving atoms in the rows 
immediately above and below them change. If we recall that bonds between neighboring atoms are only visualization aids, saying that ``one defect moves by breaking bonds between 
some atoms and creating bonds with others'' is only a figure of speech, not a consequence 
of the model.

Our models show that edge dislocations can have cores different from those depicted in 
Figure \ref{figura2}. In Fig.~\ref{figura4}, eight atoms, one with a dangling bond, form 
the core of an edge dislocation. When the dark atom moves to the left as indicated by the 
arrow and is attached to the light atom with the dangling bond, the light atom formerly 
connected to it moves to the right and one of its bonds is left dangling. The overall result is the 
one-step motion of the octagon defect to the right. 

 Figure \ref{figura5} depicts a vacancy defect which is a possible core of an edge dislocation 
dipole, as it will be shown later. Fig.~\ref{figura5}(a) is the initial configuration obtained
by discretization of the elastic field of an edge dislocation dipole. Fig.~\ref{figura5}(b) is
the final configuration obtained by evolution of (a).

\begin{figure}
\begin{center}
\includegraphics[width=8cm]{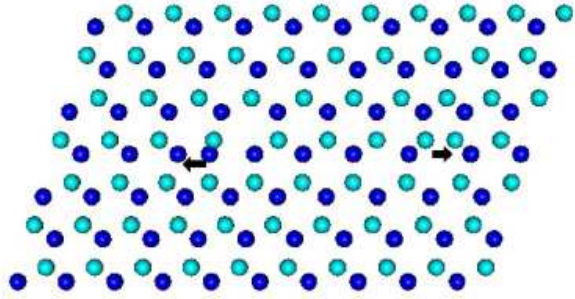}
\caption{(Color online) The two component 5-7 edge dislocations of a SW defect separate 
under a large applied shear stress. This figure has been calculated using a simple periodized 
scalar model in a lattice with $18\times 18$ lattice spacings.}
\label{moving}
\end{center}
\end{figure}

Figure \ref{figura3}(a) depicts another possible core of an edge dislocation dipole: a
divacancy formed by an octagon with two adjacent pentagons. This 5-8-5 defect is dynamically stable. Yet another different core of the dipole is the Stone-Wales 
(SW) defect formed by a pair of heptagon-pentagons, as depicted in Fig.~\ref{figura3}(b). 
This 5-7-7-5 defect is created by the SW bond rotation of one
light atom forming the lowest vertex of an hexagon and bond reattachment to another light
atom to form the top of a pentagon and, at the same time, the basis of an heptagon. This 
bond rotation leaves an oppositely oriented 5-7 defect next to the other one.
This SW defect has been obtained by superposition of the initial guesses for the elastic 
displacement fields of two opposite edge dislocations with heptagon-pentagon cores that share
the same glide line. In elasticity, we would expect that the two component edge dislocations
of this dipole glide towards each other and annihilate, leaving the undisturbed lattice as a 
result. This is in fact what the numerical solution of our model shows. Thus the SW defect 
represents an edge dislocation dipole and it seems to be dynamically unstable \cite{stable}. 
If a sufficiently large shear stress is applied to a lattice that contains one SW defect, its two
component dislocations drift apart as shown in Fig.~\ref{moving}. 

If we form a different dipole with component dislocations having opposite Burgers vectors
and different glide lines, the two component edge dislocations glide towards each other but
do not meet. Instead they form a stable dislocation dipole. As in the case of nanotubes
\cite{orl99}, the simplest realization of this idea is a 7-5-5-7 defect in which the two 
pentagons (not the two heptagons as in the SW configuration) share a common side. Our 
results show that the 7-5-5-7 defect is dynamically stable \cite{geli}.

\section{Discrete elasticity models for graphene sheets}
\label{sec:modgraf}
We want to produce a lattice model that reduces to the equations of linear isotropic
elasticity in the far field of a defect. In the continuum limit, elastic
 deformations of graphene sheets are described by the Navier equations for the 2D displacement vector $(u,v)$:
 \begin{eqnarray}
 \rho {\partial^2 u\over \partial t^2}= 
 C_{11} {\partial^2 u\over \partial x^2} +
 C_{66} {\partial^2 u\over \partial y^2} +
 ({C_{66}+C_{12}}){\partial^2 v\over \partial x \partial y},
 \label{s3}\\
 \rho {\partial^2 v\over \partial t^2}= 
 C_{66} {\partial^2 v\over \partial x^2} +
 C_{11} {\partial^2 v\over \partial y^2} +
 ({C_{66} +C_{12}}){\partial^2 u\over \partial x \partial y},
 \label{s4}
 \end{eqnarray}
 where $\rho$ is the mass density. In the basal plane, graphite is
 isotropic, so that $C_{66}=\mu$, $C_{12}= \lambda$ and $C_{11}=
 \lambda + 2\mu$, in which $\lambda$ and $\mu$ are the Lam\'e 
 coefficients. 
 
\subsection{Discretization of the Navier equations and linear lattice model}
We would like to find a linear lattice model by discretizing the Navier equations (\ref{s4}). This is not an obvious task because any symmetric combination of differences involving either all nearest neighbors, or all nearest and next nearest neighbors, etc.\ only yields a multiple of the Laplacian in the continuum limit, no matter how many neighbors we use. We have allowed combinations of differences that do not involve all neighbors of the same type symmetrically. The idea is to select three difference operators that yield three independent linear combinations of $\partial^2 u/\partial x^2$, $\partial^2 u/\partial y^2$ and $\partial^2 u/\partial x\partial y$ in the continuum limit. Then we replace the partial derivatives of $u$ in (\ref{s4}) by the combinations of our differences that provide those partial derivatives in the continuum limit. 

Let us select assign the coordinates $(x,y)$ to the atom $A$ in sublattice 1 (see Figure 
\ref{figura1}). The three nearest neighbors of $A$ belong to sublattice 2 and their cartesian coordinates 
are $n_1$, $n_2$ and $n_3$ below. Its six next-nearest neighbors belong to sublattice 1 and their cartesian coordinates are $n_i$, $i=4,\ldots, 9$:
\begin{eqnarray}
&& n_{1}=\left(x-{a\over 2},y-{a\over 2\sqrt{3}}\right), \, n_{2}=\left(x+{a\over 2},
y-{a\over 2\sqrt{3}}\right), \, n_{3}=\left(x,y+{a\over\sqrt{3}}\right),\nonumber\\
&& n_{4}=\left(x-{a\over 2},y-{a\sqrt{3}\over 2}\right), \, n_{5}=
\left(x+{a\over 2},y-{a\sqrt{3}\over 2}\right), \, n_{6}=(x-a,y), 
\nonumber\\
&& n_{7}=(x+a,y),\, n_{8}=\left(x-{a\over 2},y+{a\sqrt{3}\over 2}\right),\, 
n_{9}=\left(x+{a\over 2},y+{a\sqrt{3}\over 2}\right).\label{neiA}
\end{eqnarray}
Four of these atoms are separated from $A$ by the primitive vectors $\pm {\bf a}$ and 
$\pm {\bf b}$ (see Figure \ref{figura1}). 

\begin{figure}
\begin{center}
\includegraphics[width=8cm]{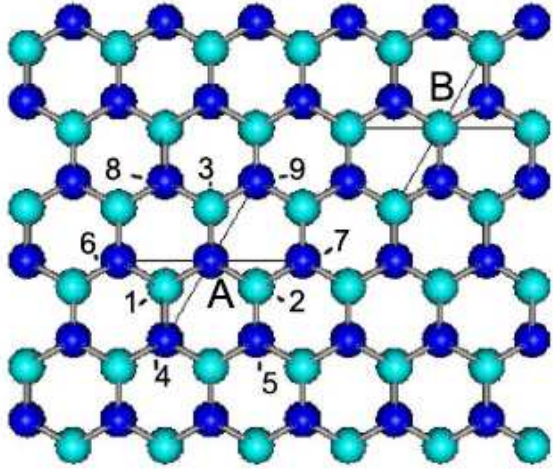}
\caption{(Color online) Neighbors of a given atom $A$. Only the neighbors labeled 1, 2, 
3, 4, 6, 7 and 9 are affected by the difference operators $T$, $H$ and $D_{1}$ used in our 
discrete elasticity model with two slip directions. }
\label{figura6}
\end{center}
\end{figure}

Let us define four operators acting on functions of the coordinates $(x,y)$ of node $A$: 
\begin{eqnarray}
Tu &=& [u(n_{1})-u(A)] + [u(n_{2})-u(A)] + [u(n_{3})-u(A)],\label{T}\\
Hu & = & [u(n_{6})-u(A)]+[u(n_{7})-u(A)],  \label{H} \\
D_{1}u &=& [u(n_{4})-u(A)] + [u(n_{9})-u(A)],  \label{D1}\\
D_{2}u &=& [u(n_{5})-u(A)] + [u(n_{8})-u(A)].  \label{D2} 
 \end{eqnarray}
Note that the operator $T$ involves finite differences with the three next neighbors of 
$A$ which belong to sublattice 2, whereas $H$ and $D_{1}$ involve differences between 
atoms belonging to the same sublattice along the primitive directions $\mathbf{a}$ 
and $\mathbf{b}$, respectively. See Figure \ref{figura6}. $D_{2}$ involves differences 
between atoms belonging to the same sublattice along a different choice of the basis vectors: 
$\mathbf{a}$ and $\mathbf{c}$ (parallel to the line joining atoms $n_{5}$ and $n_{8}$ in 
Fig.~\ref{figura6}, which is also another primitive direction). The operator $D_{2}$ will be 
important when we want to consider a dislocation motion along slip directions $\mathbf{a}$ 
and $\mathbf{c}$. Taylor expansions of these finite difference combinations about $(x,y)$ 
yield
\begin{eqnarray}
Tu &\sim& \left({\partial^2 u\over\partial x^2} + {\partial^2 u\over\partial 
y^2}\right) {a^2\over 4},\nonumber\\
Hu &\sim& {\partial^2 u\over\partial x^2}\,  a^2, \nonumber\\
D_{1}u &\sim& \left({1\over 4}\, {\partial^2 u\over\partial x^2} +{\sqrt{3}\over 2}\,
{\partial^2 u\over\partial x\partial y} + {3\over 4}\, {\partial^2 u\over
\partial y^2}\right)a^2,  \nonumber\\
D_{2}u &\sim& \left({1\over 4}\, {\partial^2 u\over\partial x^2} -{\sqrt{3}\over 2}\,
{\partial^2 u\over\partial x\partial y} + {3\over 4}\, {\partial^2 u\over
\partial y^2}\right)a^2,  \nonumber
\end{eqnarray}
as $a\to 0$. 

\subsubsection{Model with three slip directions}
Let us assume that we want to allow dislocations to slip along any of the three primitive
directions $\mathbf{a}$, $\mathbf{b}$ or $\mathbf{c}$. Then we replace in 
Equations (\ref{s3}) and (\ref{s4}) $Hu/a^2$, $(4T-H)u/a^2$ and $(D_{1}-D_{2})u/(
\sqrt{3}a^2)$ instead of $\partial^2 u/\partial x^2$, $\partial^2 u/\partial y^2$ 
and $\partial^2 u/\partial x\partial y$, respectively, with similar substitutions for the 
derivatives of $v$. In terms of the Lam\'e coefficients, we obtain the following equations at 
each point of the lattice:
\begin{eqnarray}
\rho a^2 {\partial^2 u \over \partial t^2} = 4\mu\, Tu + (\lambda+\mu)\, Hu 
+{\mu+\lambda \over\sqrt{3}}\, (D_{1} - D_{2})v, \label{co1}\\
\rho a^2 {\partial^2 v\over \partial t^2}= 4 \mu\, Tv + ( \lambda+\mu )\,
(4T- H)v +{\mu+\lambda\over\sqrt{3}}\, (D_{1} - D_{2})u. \label{co2}
 \end{eqnarray}

\subsubsection{Model with two slip directions}
If we only allow slip along the two primitive directions $\mathbf{a}$ and $\mathbf{b}$
that form our vector basis, we should replace in Equations (\ref{s3}) and (\ref{s4}) 
$Hu/a^2$, $(4T-H)u/a^2$ and $2(D_{1}-3T+H/2)u/(\sqrt{3}a^2)$ instead of $\partial^2 
u/\partial x^2$, $\partial^2 u/\partial y^2$ and $\partial^2 u/\partial x\partial 
y$, respectively, with similar substitutions for the derivatives of $v$. In terms of the Lam\'e 
coefficients, the following equations are then obtained
\begin{eqnarray}
\rho a^2 {\partial^2 u \over \partial t^2} = 4\mu\, Tu + (\lambda+\mu)\, Hu 
+{2(\mu+\lambda) \over\sqrt{3}}\left(D_{1} - 3T + {1\over 2} H\right)v, 
\label{e1}\\
\rho a^2 {\partial^2 v\over \partial t^2}= 4 \mu\, Tv + ( \lambda+\mu )\, (4T- H)v 
+{2  (\mu+\lambda) \over\sqrt{3}}\left(D_{1} - 3T + {1\over 2} H\right)u,  
\label{e2}
 \end{eqnarray}
at every point of the lattice. To have slip directions $\mathbf{a}$ and $\mathbf{c}$, we
replace the operator $-(D_{2} - 3T + H/2)$ instead of $(D_{1} - 3T + H/2)$ in (\ref{e1})
and (\ref{e2}). The same equations are found if $B=(x,y)$ is an atom in 
sublattice 2. In this case, the relevant neighbors of $B$ entering the definitions of $T$, $H$,
$D_{1}$ and $D_{2}$ have coordinates
 \begin{eqnarray}
&& n_{1}=\left(x-{a\over 2},y+{a\over 2\sqrt{3}}\right), \,
n_{2}=\left(x+{a\over 2},y+{a\over 2\sqrt{3}}\right), \, n_{3}=\left(x,y-
{a\over \sqrt{3}}\right),\nonumber\\
&& n_{4}=\left(x-{a\over 2},y-{a\sqrt{3}\over 2}\right),\,
n_{5}= \left(x+{a\over 2},y-{a\sqrt{3}\over 2}\right),\, n_{6}=(x-a,y),\nonumber\\ 
&& n_{7}=(x+a,y),\, n_{8}=\left(x-{a\over 2},y+{a\sqrt{3}\over 2}\right),\,
n_{9}=\left(x+{a\over 2},y+{a\sqrt{3}\over 2}\right). \label{neiB}
 \end{eqnarray}
 
We may see the hexagonal lattice as a set of atoms connected by springs. These
springs connect each atom $A$ with its nearest neighbors $n_{1}$, $n_{2}$ and $n_{3}$, 
and with its nearest neighbors along the primitive directions, $n_{4}$, $n_{6}$, $n_{7}$ 
and $n_{9}$. If we add symmetrically the missing two next-nearest neighbors $n_{5}$ and $n_{8}$
in the operator $D_{1}$, its Taylor expansion produces the 2D Laplacian. Similarly, adding
all the second-nearest neighbors, the Laplacian is found again: Symmetric choices of 
neighbors only generate Laplacians, but not the terms involving cross derivatives.
It seems reasonable to break the central symmetry about a given atom when defining
finite differences by giving preference to the primitive directions.
Notice that if we move along the lattice, the hexagonal arrangement
itself is a source of anisotropy. Along the $x$ direction (a primitive
direction of the lattice), atoms are arranged in a `zig-zag' pattern. The 
same arrangement occurs along the other primitive directions. However, along the $y$ 
direction atoms are arranged in an  `arm-chair' pattern. 

\subsection{Lattice model in primitive coordinates}
Equations (\ref{e1}) and (\ref{e2}) can be written in primitive coordinates 
$u'_{i}$, $i=1,2$ (with $u'_{1}=u'$, $u'_{2}=v'$), by means of the transformation $u_{i}=
\mathcal{T}_{ij}u'_{j}$ (summation over repeated indexes is intended), with
\begin{eqnarray}
\left(\begin{array}{c}
u\\
v\end{array}\right)=a \left(\begin{array}{cc}
1&{1\over 2}\\
0 & {\sqrt{3}\over 2}\end{array}\right)\left(\begin{array}{c}
u'\\
v'\end{array}\right). \label{e3}
\end{eqnarray}
Writing Equations (\ref{co1}) and (\ref{co2}) or (\ref{e1}) and (\ref{e2}) as 
$\rho a^2\partial^2 u_{i}/\partial t^2= L_{ij} u_{j}$, the {\it  linear  equations of 
motion in the primitive coordinates} are $\rho a^2\partial^2 u'_{i}/\partial t^2= L'_{ij} 
u'_{j}$ with $L'_{ij}= \mathcal{T}^{-1}_{ik} L_{kn}\mathcal{T}_{nj}$, or equivalently:
\begin{eqnarray}
\rho a^2\, {\partial^2 u' \over \partial t^2}= 4\mu\, Tu' +
{\lambda +\mu\over  3}\, (3H-D_{1}+D_{2})u' + (\lambda+\mu)\left(H+\frac{D_{1}
-D_{2}}{3}-2T\right)v', \label{co1p} \\
\rho a^2\, {\partial^2 v'\over \partial t^2}=
{\lambda+\mu\over 3}\, (D_{1}-D_{2}-3H)v' + 4(\lambda+2\mu)Tv'
+ \frac{2}{3}(\lambda+\mu) (D_{1}-D_{2})u', \label{co2p}
\end{eqnarray}
when there are three slip directions, or
\begin{eqnarray}
{\rho a^2\over 2} {\partial^2 u' \over \partial t^2}=
{\lambda +\mu\over  3}\, [(H-D_{1})u' + (2H+D_{1})v'] + T[(\lambda +3\mu)u'-
2(\lambda+\mu)v'], \label{e1p} \\
{\rho a^2\over 2} {\partial^2 v'\over \partial t^2}=
{\lambda+\mu\over 3}\, [(H+2D_{1})u' + (D_{1}-H)v'] + T[(\lambda +3\mu)v'-
2(\lambda+\mu)u'], \label{e2p}
\end{eqnarray}
when there are only two slip directions along the basis vectors $\mathbf{a}$ and 
$\mathbf{b}$. Note that $u'=(u-v/\sqrt{3})/a$ and $v' = 2v/(a\sqrt{3})$ are 
nondimensional. Equations (\ref{co1p}) and (\ref{co2p}) do not look symmetric in the 
same way as (\ref{e1p}) and (\ref{e2p}) do because we have selected a basis 
along primitive directions $\mathbf{a}$ and $\mathbf{b}$, so that $\mathbf{c}$ (which 
defines the third slip direction associated to the operator $D_{2}$) is not a basis vector.
 
 \subsection{Scalar model}
\label{sec:scalar}
When the displacements in the $y$ direction are negligible we may ignore
the vertical component and work in cartesian coordinates. The evolution 
equations for the displacements in the horizontal direction are:
 \begin{eqnarray}
\rho a^2 {\partial^2 u\over\partial t^2} = 4\mu\, Tu + (\lambda+\mu)\, Hu. 
 \label{e1r}
 \end{eqnarray}
This equation can also be obtained from $u=(u'+v'/2)\, a$ by adding twice (\ref{e1p}) to
(\ref{e2p}), and then setting $v'=0$. The only slip direction of the scalar model is along
 $\mathbf{a}$.

\section{Periodized discrete elasticity models for graphene sheets}
\label{sec:periodic}
The models described by (\ref{co1}) - (\ref{co2}), (\ref{e1}) - (\ref{e2}) or (\ref{e1r})
are linear and do not allow for the changes of neighbors involved in defect motion. An 
obvious way to achieve this is to update neighbors as a defect moves. Models such as 
(\ref{co1}) - (\ref{co2}), (\ref{e1}) - (\ref{e2}) or (\ref{e1r}) would have the same 
appearance, but the neighbors $n_i$ would be given by (\ref{neiA}) and (\ref{neiB}) only 
at the start. At each time step, we keep track of the position of the different atoms and update 
the coordinates of the $n_{i}.$ This is commonly done in Molecular Dynamics, as 
computations are actually carried out with only a certain number of neighbors. Convenient as 
updating is, its computational cost is high and analytical studies thereof are not easy. Another
advantage of periodized discrete elasticity is that boundary conditions can be controlled 
efficiently to avoid spurious numerical reflections at boundaries.

In simple geometries, we can avoid updating by introducing a periodic function of differences in the primitive directions that 
automatically describes link breakup and union associated with defect motion. Besides greatly
reducing computational cost, the resulting periodized discrete elasticity models allow 
analytical studies of defect depinning \cite{car03,dm1,dm2}, motion and nucleation 
\cite{pcb}. 

To restore crystal periodicity, we replace the linear operators $T$, $H$, $D_{1}$ and $D_{2}$ 
in (\ref{co1p}) - (\ref{co2p}) or (\ref{e1p}) - (\ref{e2p}) by their periodic versions:
\begin{eqnarray}
&& T_{p}u'=g(u'(n_{1})-u'(A)) + g(u'(n_{2})-u'(A)) + g(u'(n_{3})-u'(A)),\label{t-per}\\
&& H_{p}u'= g(u'(n_{6})-u'(A))+g(u'(n_{7})-u'(A)), \label{h-per}\\
&& D_{1p}u'=g(u(n_{4})-u'(A)) + g(u'(n_{9})-u'(A)),\label{d-per}\\
&& D_{2p}u'=g(u(n_{5})-u'(A)) + g(u'(n_{8})-u'(A)),\label{d2-per}
\end{eqnarray}
where $g$ is a periodic function, with period one, and such that $g(x)\sim x$ as $x\to 0$.
In our tests we have taken $g$ to be a periodic piecewise linear continuous function: 
\begin{eqnarray}\label{periodic}
g_{\alpha}(x)= \left\{ 
\begin{array}{ll}
x, & -\alpha \leq x \leq \alpha, \\
-{2 \alpha \over 1 - 2 \alpha}  x + {\alpha \over 1 - 2 \alpha},
  &  \alpha \leq x \leq 1 -\alpha.
\end{array}\right.
\end{eqnarray}
The parameter $\alpha$ controls defect stability and mobility under applied stress. $\alpha$ 
should be sufficiently large for elementary defects (dislocations, vacancies) to be stable at 
zero applied stress, and sufficiently small for dislocations to move under reasonable applied 
stress \cite{dm1}. The periodic function $g$ can be replaced by a different type of periodic 
function to achieve a better fit to available experimental or numerical data.

Periodized discrete elasticity is a Lagrangian model: the atoms are labeled from the start and
we track their motion. The periodic functions allow us to simulate dislocation motion 
without updating neighbors thereby greatly reducing computational cost. 

In the simpler case of scalar elasticity (\ref{e1r}), the corresponding periodized discrete 
elasticity model is:
 \begin{eqnarray}
&&\rho a^2 {\partial^2 u\over\partial t^2} = 4\mu\, T_{p}u + (\lambda+\mu)\, 
Hu, \label{e1sper}\\
&& T_{p}u = a\, g\left({u(n_{1})-u(A)\over a}\right) + a\, g\left({u(n_{2})-u(A)
\over a}\right) + a\, g\left({u(n_{3})-u(A)\over a}\right). \label{e2sper}
 \end{eqnarray}
The nonlinear function $g$ is only needed for differences between the neighbors that may change due to defect motion. In the direction $x$, neighbors never change, therefore we use the operator $H$ of Eq.\
(\ref{H}). Horizontal rows may shift, resulting in a shift of neighbors that is taken 
into account by using $T_{p}$ as in (\ref{e2sper}) with the periodic function $g_{\alpha}$ 
defined in (\ref{periodic}).

\section{Defects in graphene}
\label{sec:defgraf}
In this Section, we discuss the defects obtained with our periodized
discrete elasticity models. In our numerical calculations, we use for graphene the elastic 
constants of graphite in the basal plain (which is isotropic), $C_{11}=C_{12}+ 2C_{66}= 
1060$ GPa, $C_{12}=\lambda=180$ GPa, $C_{66}=\mu=440$ GPa.\cite{bla70} This 
yields $\nu= 0.17$. We have used $\alpha=0.2$ in the periodic function $g_{\alpha}(x)$. 
For this value, the Peierls stress for a 5-7 defect is 0.025 $\mu$, which is of the order of 
known values for covalent crystals \cite{hull}. Larger (smaller) values of $\alpha$ yield 
larger (smaller) Peierls stresses: for $\alpha$ between 0.15 and 0.3, the Peierls stress varies
in the range between $10^{-3}\mu$ and $10^{-1}\mu$. We have not calibrated our model
to a precise value but this can be done when Peierls stresses for single graphene sheets are
measured in experiments. All defects have been calculated using the 
vectorial model (\ref{e1p}) - (\ref{e2p}), periodized by means of (\ref{t-per}) -
(\ref{d-per}). Analogous results can be found using appropriate computationally cheaper scalar models such as (\ref{e1sper}) - (\ref{e2sper}). Scalar models are more
convenient to analyze defect motion, interaction and even nucleation of defects constrained to move along a given primitive direction. 
In more complex geometries a combination of periodized discrete
elasticity models and neighbor updating could be useful.

In all cases, the construction of defects is similar. We use the displacement field of a given defect in continuum linear elasticity at zero applied stress both as initial and boundary condition for the discrete model. Then the model with overdamped dynamics (replacing second order time derivatives in the model equations by 
first order ones) is used to relax the initial field to the stable stationary solutions representing the sought defects. Applied stress can be implemented via the boundary conditions. To study defect motion and interaction, we can use the models with inertia. 

How large should our computational lattice be? We can have an idea by using results by
Zhang et al \cite{zha06}. As the far field of defects, they considered linear elasticity with a 
strain energy that was a quadratic functional of the strain tensor and also of its first and 
second derivatives. They compared results given by this theory with those of classical linear 
elasticity and with results from atomistic simulations. In their figure 11, it can be seen that 
four lattice spacings away from the core of a SW defect or of a divacancy in graphene, 
linear elasticity already approximates very well MD results (they use only 132 carbon atoms 
in their simulations). Thus a lattice with $18\times 18$ 
lattice spacings (and therefore with $36\times 36= 1296$ atoms) should be sufficiently large 
to obtain good results for a centrally located defect by using our discrete elasticity models. 
The figures presented in this paper have been obtained using such a lattice and we have 
checked that our results do not change significantly by decreasing slightly the lattice size or 
by increasing it. 

As we have said before, our model is more efficient than MD. In MD, one has two options
when computing the force acting on each atom, either to include the interactions of one 
atom with all its neighbors, or to include only the neighbors within a certain distance of 
the atom and update them as time evolves. The cost of computing the force on 
each atom at each time step according to the first option is huge due to the large number 
of atoms involved, and therefore we need a very long time to compute just a few time steps.
The second option (few neighbors and updating) is less computationally costly than the first
one, but its cost is still large. In contrast to this, our model involves a few nearest neighbors 
and the periodic function used in the governing equations avoids the need of updating 
neighbors at each step. Thus computation of the forces at each step is cheap. We may 
compute the evolution of atoms in much larger lattices in a much shorter time. 
Since our forces are cheaper to compute, we may use higher order
solvers to integrate the resulting system of differential equations.
Since we are able to handle larger lattices, our trouble with numerical
artifacts due to reflections on the walls of the lattice is smaller. In MD, reflections of
waves at the boundaries of the computational domain limit the time spans over
which simulations are reliable and there is a trade off between lattice size and the total time
of a simulation. In fact, it is not yet clear how to prevent spurious reflections
in MD simulations in an efficient manner\cite{shi05}. The simplicity of our model 
equations allows us to introduce simple non reflecting boundary conditions, thereby 
suppressing numerical size effects due to reflections at the boundaries. 

\subsection{Edge dislocations}
Static edge dislocations can be generated using the overdamped version of 
(\ref{e1p})-(\ref{e2p}) periodized by means of (\ref{t-per}) - (\ref{d-per})
and the elastic field of edge dislocations for (\ref{e1})-(\ref{e2}).
To find the stationary edge dislocation at zero stress, we first write the 
corresponding stationary edge dislocation of isotropic continuum elasticity. The displacement 
vector ${\bf u}= (u(x,y),v(x,y),0)$ of an edge dislocation directed along the $z$ axis 
(dislocation line), and having Burgers vector $(b,0,0)$ is 
\begin{eqnarray} 
u &= & {b \over 2 \pi} \left[\tan^{-1}\left({y\over x}\right) + 
{ xy \over 2(1-\nu)(x^2+y^2)}\right],\nonumber\\  
v &= & {b \over 2 \pi} \left[ -{1-2\nu\over 4(1-\nu)}\,\ln\left( 
{x^2+y^2\over b^2}\right) + {y^2 \over 2(1-\nu) (x^2+y^2)}\right], \label{dy4}
\end{eqnarray}
cf.\ Ref.~\onlinecite{nab67}, pag.\ 57. (\ref{dy4}) has
a singularity $\propto r^{-1}$ at the core and it satisfies $\int_{\mathcal{ C}} 
(d{\bf x}\cdot\nabla) {\bf u} =- (b,0,0)$, for any closed curve $\mathcal{C}$ 
encircling the $z$ axis. It is a solution of the planar stationary Navier equations with a singular 
source term:
\begin{eqnarray}
\Delta {\bf u} + {1\over 1-2\nu} \nabla(\nabla\cdot {\bf u}) = -(0,b,0)\,\delta(r).  
  \label{dy3}
\end{eqnarray}
Here $r=\sqrt{x^2+y^2}$ and $\nu=\lambda/[2(\lambda + \mu)]$ is the Poisson ratio; 
cf.\ page 114 of Ref.~\onlinecite{ll7}. 

The continuum displacement (\ref{dy4}) yields the nondimensional static displacement vector in primitive coordinates $U'(l,m)= 
[u(x,y)-v(x,y)/\sqrt{3}]/a$, $V'(l,m)=2v(x,y)/(a\sqrt{3})$, where $x=a(x'+y'/2)$, $y=
a\sqrt{3}\, y'/2$. The primitive coordinates $x'=x'_0 + l$, 
$y'=y'_0+m$ are centered in an appropriate point $(x'_0,y'_0)$  which is different from the origin to avoid the singularity in (\ref{dy4}) to coincide with a lattice point. 

${\bf U}'(l,m)$ will be used to find the 
stationary edge dislocation of the discrete equations of motion. 
To this end, we
replace the inertial terms $\rho a^2\partial^2 u'/\partial t^2$ and $\rho a^2\partial^2 v'/\partial t^2$ in (\ref{e1p}) - (\ref{e2p}) by $\beta\,\partial u'/\partial t$ and $\beta\,\partial v'/\partial t$, respectively. The 
resulting overdamped equations have the same stationary solutions. We use
an initial condition ${\bf u'}(l,m;0) = {\bf U'}(l,m)$ given 
by Eqs.\ (\ref{dy4}), and boundary conditions ${\bf u'}(l,m;t) = {\bf U'}(l,m)+ (F\, 
m,0)$ at the lattice boundaries ($F$ is a dimensionless applied shear stress). 
If $|F|<F_{cs}$ ($F_{cs}$ is the static Peierls stress for edge dislocations), the 
solution   relaxes to a static edge dislocation 
$(u'(l,m),v'(l,m))$ with the appropriate continuum far field.

Depending on the location of the singularity $(x'_0,y'_0)$, there are two possible 
configurations corresponding to the same edge dislocation in the continuum limit. Figure 
\ref{figura2} shows the structure adopted by the deformed lattice $(l+u'(l,m),m+v'(l,m))$ 
when the singularity is placed between two atoms that form any non-vertical side of a given 
hexagon. The core of the dislocation is a 5-7 defect. If the singularity is 
placed in any other location different from a lattice point, the core of the singularity forms an 
octagon having one atom with a dangling bond, as shown in Fig.~\ref{figura4}. The
dangling bond causes this configuration to be more reactive due to the possibility of attaching 
impurity atoms to the dangling bond. The octagon can also be seen as a 5-7 defect with an
extra atom inserted between heptagon and pentagon. The basin of attraction of the octagon 
configuration seems to be larger than that of the 5-7 defect. That the same dislocation type 
may have two different cores is a familiar fact in crystals with diamond structure and covalent 
bonds, such as silicon; see page 376 in Ref.~\onlinecite{hirth}. There it is shown that the 
$60^\circ$ edge dislocations may belong to the `glide set' or to the `shuffle set'. Seen from 
a certain direction, the cores of the glide set look like 5-7 defects whereas the cores of the 
shuffle set look like octagons with a dangling bond attached to one of their atoms. Ewels
et al use the names ``glide dislocations'' and ``shuffle dislocations'' for the 5-7 defects and the 
octagons with a dangling bond (glide + adatom), respectively \cite{EHB02}.

The glide motion of edge dislocations occurs in the direction of their Burgers vector and on
the glide plane defined by the Burgers vector and the dislocation line. In the configurations 
of Figs.~\ref{figura2} and \ref{figura4}, a supercritical applied shear stress will move the 
dislocation in the $x$ direction on the glide plane $xz$. Our simulations show that the shuffle
dislocations move more easily than the glide dislocations, as predicted by Ewels et al 
\cite{EHB02}. For conservative or damped 
dynamics, the applied shear stress has to surpass the static Peierls stress to depin a static 
dislocation, and a moving dislocation propagates provided the applied stress is larger than 
the dynamic Peierls stress (smaller than the static one) \cite{car03}. 
A moving dislocation is a discrete traveling wave advancing along the $x$ axis, 
and having far field $(u'(l-ct,m)+Fm,v'(l-ct,m))$. The analysis of depinning and 
motion of planar edge dislocations follows that explained in Ref.~\onlinecite{car03} with 
technical complications due to our more complex discrete model. 

Similar results are obtained with the simpler scalar model (\ref{e1sper}) - (\ref{e2sper}). In this case, the continuum displacement vector of an edge dislocation at zero applied stress is 
$$U(x,y)={b \over 2 \pi}\, \theta\left(x-x_{0},{(y-y_{0})
\sqrt{\mu}\over \sqrt{\lambda+2\mu}}\right),$$ 
where $b(1,0)$ is the Burgers
vector and $\theta(x,y)=\arctan (y/x)$. Choosing the singularity point $(x_{0},y_{0})$ 
as explained above, we obtain configurations similar to those 
in Figures \ref{figura2} and \ref{figura4} except that
there is no displacement in the vertical direction. To see the effect of
a shear stress applied in the $x$ direction, we select as initial and boundary
condition 
$$U(x,y)={b \over 2 \pi}\, \theta\left((x-x_{0}),{(y-y_{0})
\sqrt{\mu}\over \sqrt{\lambda+2\mu}}\right) + F y.$$ 

\subsection{Edge dislocation dipoles}
An edge dislocation dipole is formed by two edge dislocations with
Burgers vectors in opposite directions. Depending on how we place 
the cores of these dislocations, different dipole configurations result.
Let ${\bf E}(x,y)$ be the displacement vector (\ref{dy4}) corresponding to the edge dislocation we have considered before. 
The static configuration corresponding to a dipole is either a vacancy (Fig.~\ref{figura5}),
a divacancy (Fig.~\ref{figura3}a) or a Stone-Wales defect (Fig.~\ref{figura3}b). To 
obtain these different dipole cores, we use the following initial and boundary
conditions at zero stress:
\begin{itemize}
\item Vacancy ${\bf E}(x-x_{0},y-y_{0}-l/2)-{\bf E}(x-x_{0},y-y_{0})$.
$l=a/\sqrt{3}$ is the hexagon side in terms of the lattice constant $a$.
\item Divacancy ${\bf E}(x-x_{0},y-y_{0}-l)-{\bf E}(x-x_{0},y-y_{0})$. 
\item Stone-Wales: ${\bf E}(x-x_{0}-a,y-y_{0})-{\bf E}(x-x_{0},y-y_{0})$.
\end{itemize}
We have set $x_{0}=-0.25\, a$, $y_0=-0.4\, l$ to draw Figures \ref{figura5} 
and \ref{figura3}. 

The vacancy represented in Fig.~\ref{figura5}(b) is similar to that proposed in 
Ref.~\onlinecite{kel98} using tight-binding calculations in graphite surfaces. The 
initial configuration represented in Fig.~\ref{figura5}(a) (asymmetric vacancy) evolves
toward the dynamically stable symmetric vacancy represented in Fig.~\ref{figura5}(b)
for overdamped dynamics. For conservative dynamics and zero initial velocity, the initial 
asymmetric vacancy evolves towards a stable oscillation about the symmetric vacancy.
The symmetric vacancy has the threefold symmetry observed in experiments. In a recent 
paper, Telling et al.\cite{tel03} propose that asymmetric vacancies are stable single vacancy 
defects in graphite sheets provided atoms are allowed to be displaced from the plane (see 
Figure 1 in Ref.~\onlinecite{nor03}). They further propose that, at room temperature, the 
displaced atom rotates around the vacancy center which would also explain the threefold 
symmetry observed in experiments (and possessed by the symmetric vacancy). For the initial
conditions we have considered with either conservative or overdamped dynamics and at zero
temperature, we have not found stable configurations resembling Fig.~\ref{figura5}(a) or 
oscillations between similarly asymmetric configurations with different orientations. Thus the 
asymmetric vacancy seems to be unstable for both overdamped and conservative dynamics in 
our model \cite{stable}. If this is indeed the case, this configuration is a saddle point with the 
conservative dynamics. Even if we allow bending modes of the graphene sheet, the instability 
of this saddle is likely to persist for small bending modulus. Future work will be devoted to 
study this problem and whether our stability results for different defects change if ripples in 
graphene are allowed. 

Both vacancies and divacancies are dynamically stable in an unstressed lattice. A
sufficiently large applied stress along their glide direction (in both cases, the critical stress
is about 0.12$\mu$ for $\alpha=0.2$) splits these dipoles, 
thereby originating two edge dislocations with
opposite Burgers vectors that move in opposite directions. 

Another possible core of an edge dislocation dipole is a 5-7-7-5 SW defect, as in 
Fig.~\ref{figura3}(b). When introduced as initial condition of Eqs.~(\ref{e1p}) - 
(\ref{e2p}) periodized with (\ref{t-per}) - (\ref{d-per}), the SW configuration is likely to
be unstable under zero applied stress. In fact, this configuration corresponds to two identical 
edge dislocations that have opposite Burgers vectors and share the same glide line. The 
dislocations comprising the dipole attract each other and are annihilated, leaving an 
undistorted lattice as the final configuration. Thus SW defects seem to be dynamically 
unstable \cite{stable} except if we add external forces that produce the necessary bond 
rotation and stabilize them. If a shear stress is applied in their glide direction, these defects 
either continue destroying themselves or, for large enough applied stress (0.15$\mu$ for 
$\alpha=0.2$), are split in their two component heptagon-pentagon defects that 
move in opposite directions as shown in Fig.~\ref{moving}. Note that stability of SW
defects may be very different in small-radius nanotubes which, unlike graphene, do not have 
edges on their lateral surface and have a large curvature. 

If we shift the 5-7 defects
comprising a SW in such a way that their glide lines are not the same, then we can obtain
a configuration which is stable at zero applied stress. The simplest such case is a 7-5-5-7 
defect in which the two pentagons share one side. We have checked that this defect is 
dynamically stable \cite{geli} and that it splits in its two component edge dislocations 
when a shear stress exceeding $0.09Ê\,\mu$ is applied to the lattice for $\alpha=0.2$. 
Note that in all cases (vacancies, divacancies, 5-7-7-5 and 7-5-5-7 defects) the critical
shear stress needed to split the dipole is larger than the Peierls stress for an edge dislocation.
The reason is that splitting a dipole requires overcoming the resistance of the lattice to motion
(Peierls stress) and the attraction experienced by two edge dislocations of opposite Burgers 
vectors. The latter depends on the defect configuration which thus determines the critical
dipole splitting shear stress.

Instead of a dislocation dipole, our initial configuration may be a dislocation loop, in
which two edge dislocations with opposite Burgers vectors are 
displaced vertically by one hexagon side: ${\bf E}(x-x_{0}-a,y-y_{0})-{\bf E}(x-x_{0},y-y_{0}-l)$ ($l=a/\sqrt{3}$ is the length of the hexagon side). In principle, the dislocation loop 
could evolve to an inverse SW defect (7-5-5-7). Instead, this 
initial configuration seems to evolve towards a single octagon. If we displace
the edge dislocations vertically by $l/2$, 
${\bf E}(x-x_{0}-a,y-y_{0})-{\bf E}(x-x_{0},y-y_{0}-l/2)$, 
the resulting dislocation loop seems to evolve towards a single heptagon defect.

\subsection{Energetics}
We can have an idea about the energy associated to each defect by using the energy of the 
scalar model (\ref{e1r}) (measured with respect to the stress-free undistorted lattice), 
\begin{eqnarray}
{\cal E}&=& \sum_{n}\{ 2\mu\, \left([u(n_{1})- u(n)]^2 + [u(n_{2})- 
u(n)]^2 + [u(n_{3})- u(n)]^2 \right)\nonumber\\
&+& \frac{1}{2}(\lambda+\mu)\,\left([u(n_{6})- u(n)]^2 + [u(n_{7})- u(n)]^2\right)
\},\label{energy}
\end{eqnarray}
where we sum over all points in the hexagonal lattice, belonging to sublattices $A$ or $B$ in 
Fig.~\ref{figura6}, with neighbors given by (\ref{neiA}) or (\ref{neiB}), respectively. 
It can be seen that Eq.\ (\ref{e1r}) is equivalent to $\rho a^2 \partial^2u/\partial t^2= 
- \partial {\cal E}/\partial u$. For a lattice with $16\times 16$ lattice spacings, the 
energies associated to the defects we have described are 0.658 eV for the apparently unstable 
5-7-7-5 SW defect, 2.283 eV for the octagon with a dangling bond (shuffle dislocation), 
4.917 eV for the 5-7 defect, and in the case of the dislocation dipoles: 8.825 eV for the 
vacancy, 12.074 eV for the 5-8-5 divacancy and 9.483 eV for the stable 7-5-5-7 defect 
\cite{orl99,geli}. These values are similar to those found by Ewels et al \cite{EHB02} for 
the activation barrier to form a glide dislocation dipole (8.99 eV) and a shuffle dislocation 
(2.29 eV) in graphene; cf.\ their Figure 3. Except for the SW defect, all other dislocation 
and dislocation dipole cores are stable and are obtained by dynamical evolution using the 
governing equations of the model from the class of initial conditions we mentioned above. 

\section{Conclusions}
\label{sec:conclusions}
We have studied edge dislocations and dislocation dipoles in planar graphene at zero 
temperature by means of periodized discrete elasticity models which seamlessly match the
elastic field of dislocations and dipoles as the distance from their core increases. The cores of 
edge dislocations may be the well-known pentagon-heptagon defects of Fig.~\ref{figura2} 
(glide dislocations) or octagons with a dangling bond (shuffle dislocations \cite{EHB02}) 
as in Fig.~\ref{figura4}, depending on how we choose the initial configuration. Similarly, 
different cores are possible for edge dislocation dipoles: vacancies, 5-8-5 divacancies, 
Stone-Wales defects and 7-5-5-7 defects. Of these possible cores, symmetric vacancies,
divacancies and 7-5-5-7 defects are dynamically stable whereas asymmetric vacancies and 
5-7-7-5 SW defects are likely to be unstable. Our results show that regularizing linear 
elasticity near dislocation cores by periodized discrete elasticity is a good alternative to 
computationally intensive atomistic simulations provided defects are sparse.

\acknowledgments
We thank M.A.H. Vozmediano for useful conversations. This 
research was supported by the Spanish MECD grants MAT2005-05730-C02-01 and 
MAT2005-05730-C02-02, by the Autonomous Region of Madrid under grants
S-0505/ENE/0229 (COMLIMAMS) and CM-910143, and by PR27/05-13939.


\begin{thebibliography}{9}
\bibitem{stable}
Here stability means asymptotic stability if the dynamics of the model is overdamped
(an appropriate class of initial conditions relaxes to the defects we describe), and Lyapunov 
stability (the distance to the defect configuration does not surpass a certain value for 
appropriate class of initial conditions and for all times, similar to Lyapunov stability of the 
stable equilibrium of a frictionless pendulum) if the dynamics of the model is conservative. 
It is easy to check numerically that a configuration is stable by evolving towards it by means
of overdamped dynamics. Once we have found such a stable configuration, it is easy to check
numerically that it is stable using conservative dynamics. To say that a certain configuration
is dynamically unstable because we have not been able to reach it using overdamped dynamics 
from a given initial condition that resembles the configuration is always risky: we could have 
a stable configuration with a very narrow basin of attraction, so narrow that a small 
uncertainty in the initial condition drives the system away from it. Thus in the case of 
unstable configurations checked by the type of numerical solutions we have carried out, we 
should say: ``it seems that this configuration is dynamically unstable'' or ``this configuration 
is likely to be dynamically unstable''.
\end{thebibliography}
\end{document}